\documentclass[preprint,12pt]{elsarticle}
\usepackage{amssymb}
\usepackage{amsmath}
\usepackage{bbold}
\usepackage[hidelinks,colorlinks=true, linkcolor=blue, citecolor=blue, urlcolor=blue]{hyperref}

\journal{Journal of Statistical Mechanics : theory and experiment}

\begin{document}

\begin{frontmatter}

\title{Exact Analytical Results for the 1D Ising Chain with Periodic
Impurity Fields}

\author{M. Telfah\corref{cor1}}
\author{A. Obeidat} 

\address{Department of Physics, Jordan University of Science and Technology, Irbid, Jordan}

\cortext[cor1]{E-mail: mmtelfah24@sci.just.edu.jo  (Malek Telfah)}

\begin{abstract}
We present an exact analytical solution for the one-dimensional Ising model in the presence of an external magnetic field applied periodically to every $k$-th site. The problem is handled using the symmetrized transfer matrix approach, we derive a compact closed-form expression for the system's eigenvalues for arbitrary period $k$. From the resulting free energy, we obtain exact expressions for the magnetization and zero-field susceptibility. Explicit results are presented for $k = 1$, $k = 2$, and $k = 3$ which is considered a novel result. We further analyze the spin-spin correlation functions, deriving the correlation length and the set of position-dependent correlation strength prefactors, $A_{ij}$. The framework highlights how impurity spacing suppresses thermodynamic responses, with susceptibility scaling as $\chi \sim \beta / k$ for large $k$, offering insights into diluted magnetic systems and serving as a benchmark for quasiperiodic modulations. The correlation strengths exhibit a strong anisotropy, revealing a complex, non-local structure of spin fluctuations. These results provide a complete and rigorous benchmark for understanding the effects of periodic modulation in 1D systems.
\end{abstract}


\end{frontmatter}

\section{Introduction}
\label{sec:introduction}

The one-dimensional (1D) Ising model is one of the most fundamental examples in statistical physics for understanding interacting spin systems. It describes a chain of classical spins interacting with their neighbors, and its exact solution was first achieved by Ernst Ising a century ago \cite{Ising1925}. In its simplest form—a ring of $N$ spins $\sigma_{i}=\pm1$ with nearest‐neighbour coupling $J$ and a uniform longitudinal field $h$—it is exactly solvable via the transfer-matrix method \cite{Kramers}. This method reduces the problem of summing over all configurations to the simpler algebraic problem of finding the largest eigenvalue of a small matrix \cite{Baxter}. While the 1D model famously has no finite-temperature phase transition, its exact solvability provides a rich testing ground for fundamental concepts of statistical mechanics \cite{Domb}.

Despite this simplicity, the 1D Ising chain remains an important starting point for extensions that incorporate spatial inhomogeneity, modulated external fields, or impurity sites. The study of such impurities is a vast field, as they can fundamentally alter a system's behavior \cite{Vojta2004}. Disorder can affect the stability of symmetry-broken low-temperature phases \cite{Vojta2018}, and even a single defect can, in some cases, destroy the properties of the system \cite{Zad}. Even in quantum analogs, such as the transverse-field Ising model, impurities are known to tune critical fluctuations \cite{Huang}. Such generalizations provide solvable reference points beyond the homogeneous case and show how deviations from uniformity affect a system's thermodynamic and correlation properties. These modified models bridge the gap between perfect crystalline order and complete disorder, and have connections to a wide range of topics, including random-field models \cite{Imry, Fisher} and quasiperiodic systems \cite{Aubry}. While the homogeneous ($k=1$) and dimerized ($k=2$) chains are well-understood problems \cite{Plischke}, a general and consistent framework for arbitrary $k$ has been less clear, underscoring the need for a complete and rigorous solution for general $k$.

In the present paper, we provide this complete and rigorous solution. We employ the symmetrized transfer matrix formalism to derive the exact, general eigenvalues for an arbitrary period $k$. From this, we obtain the general free energy, magnetization, and zero-field susceptibility. We present the well-known results for $k=1$ and $k=2$ as a verification of our general method, and provide the first explicit closed-form solutions for the trimerized ($k=3$) case. We then provide a detailed analysis of the spin-spin correlation functions, deriving the correlation length and the full set of position-dependent correlation strength prefactors, $A_{ij}$. Our analysis reveals a rich physical behavior, including a non-monotonic correlation length in the presence of a field and a strong anisotropy in the correlation strengths. These exact results provide a complete benchmark for the physics of periodic modulation in 1D spin systems.

Moreover, such models of periodic modulation are not just of theoretical interest. They have direct relevance to real experimental conditions. In ultracold atomic systems, for example, optical superlattices and site-selective addressing techniques allow for the engineering of systems where every $k$-th site experiences a different local potential, making the present model directly realizable \cite{Jaksch, Bloch}. Similarly, in spintronics and materials science, periodic nanostructure engineering is a practical method for tuning magnetic anisotropy in nanomagnetic arrays \cite{Cowburn}.

\section{Methodology}
\label{Methodology}

In this section, we outline the transfer matrix approach used to solve the model \cite{Kramers, Baxter}. We begin by defining the mathematical framework, including the Hamiltonian, and then derive the transfer matrices and their eigenvalues. The partition function and thermodynamic quantities derived from these eigenvalues are discussed in the subsequent section.

\subsection{Mathematical Model}
\label{Mathematical Model}

We consider a one-dimensional chain of $N$ Ising spins $\sigma_i = \pm 1$ ($i = 1, \dots, N$) with periodic boundary conditions ($\sigma_{N+1} = \sigma_1$). The Hamiltonian is given by \cite{Ising1925}:
\begin{equation}
\mathcal{H} = -J \sum_{i=1}^N \sigma_i \sigma_{i+1} - h \sum_{j \in I} \sigma_j,
\end{equation}
where $J > 0$ is the ferromagnetic nearest-neighbor coupling constant, $h$ is the external magnetic field strength, and $I$ denotes the set of impurity sites where the field is applied. These impurity sites are periodically spaced every $k$ lattice spacings, such that the field acts only on sites $j = mk + 1$ for integer $m$. This setup interpolates between the uniform field case ($k=1$) and dilute impurity limits (large $k$).

Symmetrizing the Hamiltonian by splitting the field between the two bonds around each $i^{th}$ site: 

\begin{equation}
\mathcal{H} = -\sum_{i=1}^N\left[ J\sigma_i \sigma_{i+1} + \frac{h_i}{2}\sigma_i + \frac{h_{i+1}}{2}\sigma_{i+1} \right]
\end{equation}

Where $h_j$ = h for an impurity site $\left(j \in I\right)$, and zero otherwise. The partition function is
\begin{equation}
Z = \sum_{\{\sigma\}} \exp(-\beta \mathcal{H})
\label{eq:Z}
\end{equation}
where $\beta = 1/(k_B T)$ is the inverse temperature, and the sum is over all $2^N$ spin configurations. To solve this exactly, we employ the transfer matrix method, which exploits the chain's linear structure and periodicity.

\subsubsection{Transfer Matrices and the Eigenvalue Problem}
\label{Transfer Matrices}

The partition function Z (eq.\ref{eq:Z}) can be written as the trace of a product of transfer matrices, where each matrix carries the statistical weight from one site to the next. Defining the matrix element at site \textit{i}:

\begin{equation}
    \langle \sigma_i | T_i| \sigma_{i+1}\rangle = \exp\left( \beta J\sigma_i \sigma_{i+1} + \beta \frac{h_i}{2}\sigma_i + \beta \frac{h_{i+1}}{2}\sigma_{i+1} \right)
\end{equation}

Because $\sigma_i$ are Ising spins, $T_i$ is a 2x2 matrix:
\begin{equation}
T_i = \begin{pmatrix}
e^{\beta J} e^{\beta \frac{h_i + h_{i+1}}{2}} & e^{-\beta J} e^{\beta \frac{h_i - h_{i+1}}{2}} \\
e^{-\beta J} e^{-\beta \frac{h_i - h_{i+1}}{2}} & e^{\beta J} e^{-\beta \frac{h_i + h_{i+1}}{2}}
\end{pmatrix}
 = D(\beta \frac{h_i}{2}) T_0 D(\beta \frac{h_i}{2})
\end{equation}

Here we use:

\begin{equation}
T_0 = \begin{pmatrix}
e^{\beta J} & e^{-\beta J} \\
e^{-\beta J} & e^{\beta J}
\end{pmatrix} ,\qquad
D(x) = \begin{pmatrix}
e^{x} & 0 \\
0 & e^{-x}
\end{pmatrix}
\end{equation}

Notice that $D(0) = \mathbb{1}$ for zero field and $D(t/2) = diag ( e^{t/2}, e^{-t/2})$ for impurity site $(t = \beta h )$.\\
Given the periodicity, we group the chain into unit cells of $k$ sites, each starting with an impurity site followed by $k-1$ non-impurity sites. The transfer matrix for one unit cell is then
\begin{equation}
P_k = D(t/2) T_0^k D(t/2)
\end{equation}
For a chain with $N = M k$ sites, the partition function becomes $Z = \mathrm{Tr}(P_k^M)$.\\

To evaluate $P_k$, we first require a general expression for the n-th power of the non-field matrix $T_0$. Defining $a = e^{\beta J}$, $b = e^{-\beta J}$:

\begin{equation*}
T_0 = \begin{pmatrix}
a & b \\
b & a
\end{pmatrix}\qquad
\text{with eigenvalues} \quad
\lambda_0^\pm = a \pm b
\end{equation*}

This is symmetric and diagonalizable, so we can use $T_0^n = U \Lambda^n U^{-1}$ where $U = \frac{1}{\sqrt{2}}\begin{pmatrix}
1 & 1 \\
1 & -1
\end{pmatrix}$ is a unitary matrix of the eigenvectors of $T_0$, and the result is:

\begin{equation}
T_0^n =\begin{pmatrix}
u_n & v_n \\
v_n & u_n
\end{pmatrix}\quad
\text{where}\quad
u_n = \frac{(a+b)^n + (a-b)^n}{2}, v_n = \frac{(a+b)^n - (a-b)^n}{2}
\end{equation}

So $P_k$ is given by:

\begin{equation}
P_k =\begin{pmatrix}
u_k e^t & v_k e^t \\
u_k e^{-t} & v_k e^{-t}
\end{pmatrix}
\end{equation}

The characteristic equation $\det(P_k - \lambda I) = 0$ then gives the eigenvalues
\begin{equation}
\lambda_k^\pm = u_k \cosh(\beta h) \pm \sqrt{ u_k^2 \sinh^2 (\beta h) + v_k^2 }.
\end{equation}

These eigenvalues fully determine the partition function in the thermodynamic limit, from which thermodynamic quantities follow.

\section{Thermodynamic quantities}
\label{Thermodynamic quantities}

The thermodynamic quantities are derived in the usual way from the partition function which is now written as $Z = \mathrm{Tr}(P_k^M) = (\lambda_k^+)^M + (\lambda_k^-)^M$. In the thermodynamic limit ($M \to \infty$) larger eigenvalue dominates and the partition function is simply $Z \approx (\lambda_k^+)^M$.

Starting with the free energy per site is $f = \frac{-1}{N \beta} \ln(Z)$ : 

\begin{equation}
    f =  \frac{-1}{k\beta} \ln(\lambda_k^+)
\end{equation}

The average magnetization per site (over the entire chain) is given by:

\begin{equation*}
    m = -\frac{\partial}{\partial h} f =  \frac{1}{k \beta}\frac{\partial}{\partial h} \ln (\lambda_k^+)
\end{equation*}\\

\begin{equation}
    m = \frac{ u_k \sinh(\beta h)}{k \sqrt{ u_k^2 \sinh^2 (\beta h) + v_k^2 }}
\end{equation}\\

And the zero-field susceptibility $\chi(0) = \frac{\partial m}{\partial h}|_{h=0}$ :\\

\begin{equation}
    \chi= \frac{\beta u_k}{k v_k} = \frac{\beta}{k} \left (\frac{1+\tanh^{k}(\beta J)}{1-\tanh^k(\beta J)} \right )
\end{equation}\\

With the definitions of $u_k$ and $v_k$ given in \ref{Transfer Matrices}.\\
These are the exact solutions for general spacing k derived analytically.\\
For large k, corresponding to the case of diluted fields, $\tanh^{k} \to 0$  and the susceptibility obeys Curie - like law:

\begin{equation*}
    \chi \sim \frac{\beta}{k}
\end{equation*}

\section{Exact Solutions for Special Cases}
\label{Exact solutions for special cases}

Having derived the general expressions for the magnetization $m$ and zero-field susceptibility $\chi(0)$ in terms of $u_k$ and $v_k$, we now specialize to small periods $k=1,2,3$. These cases serve as benchmarks: $k=1$ recovers the uniform-field 1D Ising model, $k=2$ models a dimerized chain with alternating impurities, and $k=3$ provides the first explicit closed-form results for a trimerized configuration, highlighting the framework's utility for higher periods. For each, we compute $u_k$ and $v_k$ explicitly and substitute into the general formulas, yielding simplified expressions.

\subsection{Case $k=1$: Uniform Field}

For $k=1$, every site experiences the field, so $u_1=a$, $v_1=b$. Substituting into the general magnetization formula gives
\begin{equation}
m = \frac{\sinh(\beta h)}{\sqrt{\sinh^2(\beta h) + e^{-4\beta J}}},
\end{equation}
which matches the standard 1D Ising result under uniform field. The zero-field susceptibility simplifies to
\begin{equation}
\chi(0) = \beta e^{2\beta J}.
\end{equation}

This case exhibits the strongest response, with $\chi(0)$ diverging exponentially as $T \to 0$, reflecting cooperative alignment.

\subsection{Case $k=2$: Dimerized Chain}

For $k=2$, impurities alternate, yielding  $u_2=a^2 + b^2$, $v_2=2ab = 2$. The magnetization becomes
\begin{equation}
m = \frac{\sinh(\beta h)}{2 \sqrt{\sinh^2(\beta h ) + \text{sech}^2(2 \beta J)}},
\end{equation}
and the susceptibility is
\begin{equation}
\chi(0) = \frac{\beta}{2}\cosh (2 \beta J).
\end{equation}

This represents slower divergence than the exponential in k = 1 case.

\subsection{Case $k=3$: Trimerized Chain}

For $k=3$, impurities occur every third site, with $u_3=a(a^2 + 3b^2)$, $v_3=b(3a^2 + b^2)$. Substituting yields the novel closed-form magnetization
\begin{equation}
m = \frac{\sinh(\beta h)}{\sqrt{\sinh^2(\beta h) + e^{-12\beta J} \left( \frac{1 + 3 e^{4 \beta J}}{1+3e^{-4 \beta J}}  \right)^2}},
\end{equation}
and susceptibility
\begin{equation}
\chi(0) = \frac{\beta}{3}e^{6\beta J} \left(\frac{1 + 3 e^{- 4 \beta J}}{1 + 3 e^{ 4 \beta J}}\right).
\end{equation}
This expression for $k=3$ is new and demonstrates further suppression. However, for weak \textit{J} or equivalently at high \textit{T}, the expressions for k = 2 and k = 3 tend to that for k = 1 ($X(0) \sim \beta e^{2\beta J}$) except for the additional factor of 1/k reflecting the fact that \textit{m} represents the average magnetization over all the chain not only over the impurity sites $(m_{imp})$

\begin{figure}[h]
\centering
\includegraphics[width=0.8\textwidth]{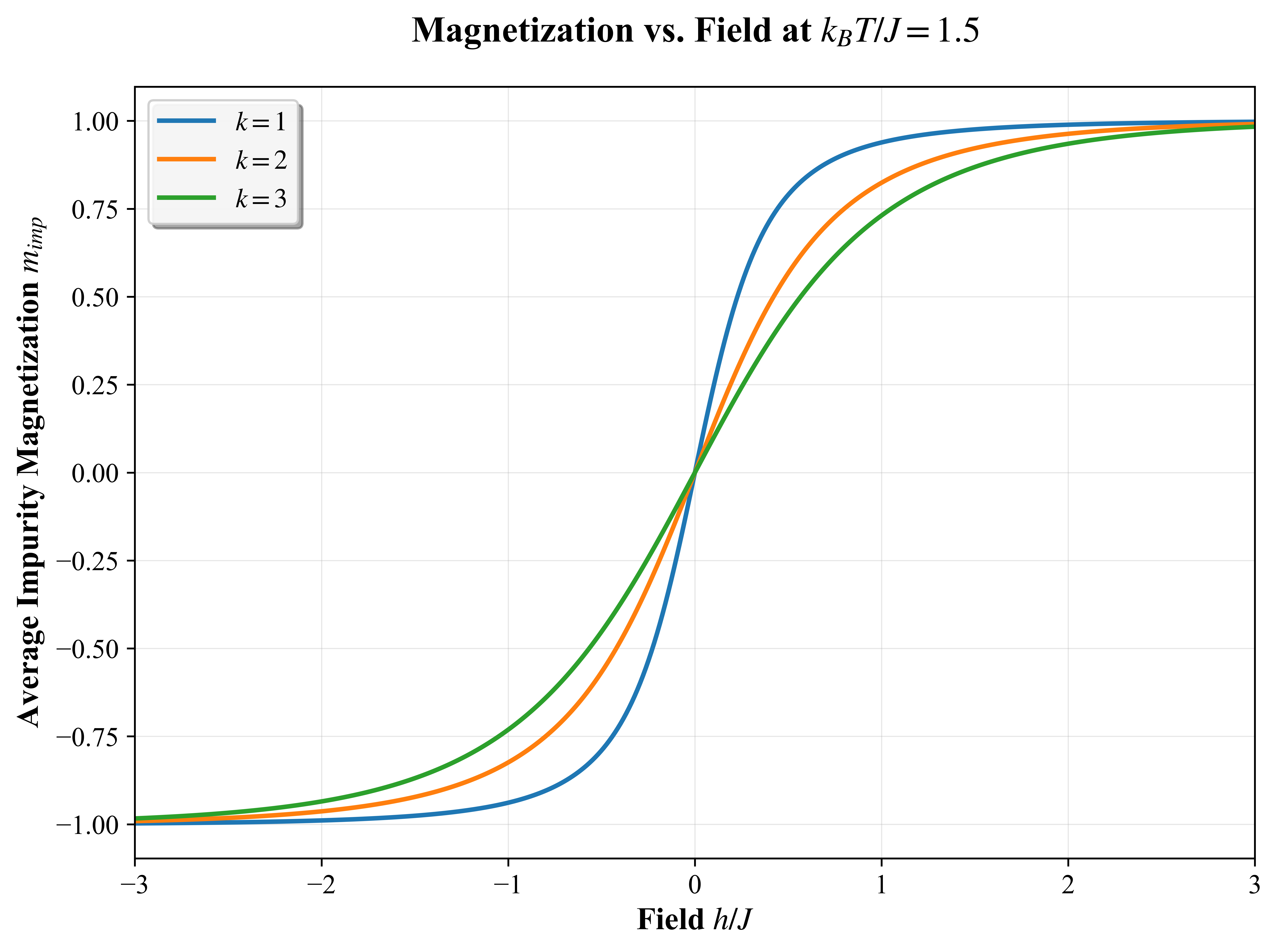}
\caption{Magnetization $m$ as a function of field $h$ at $\beta J = 1.5$, for $k=1,2,3$.}
\label{fig:m_vs_h}
\end{figure}

\begin{figure}[h]
\centering
\includegraphics[width=0.8\textwidth]{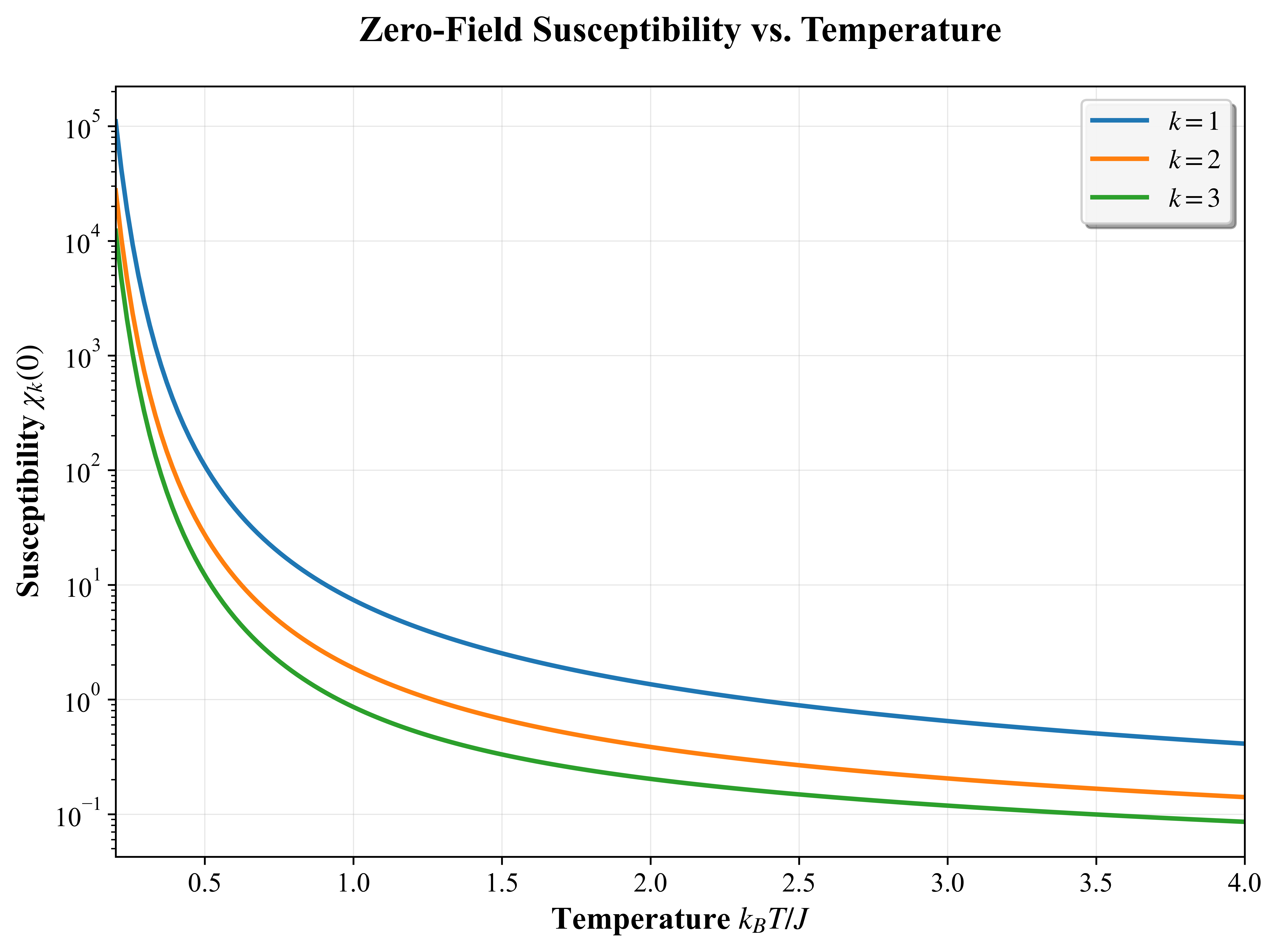}
\caption{Zero-field susceptibility $\chi(0)$ as a function of temperature $T$, for $k=1,2,3$.}
\label{fig:chi_vs_T}
\end{figure}

To visualize these results, Figure~\ref{fig:m_vs_h} shows $m_{imp}$ versus $h$ at fixed intermediate temperature ($\beta J = 1.5$), where larger $k$ reduces the slope near $h=0$ and saturation value. Figure~\ref{fig:chi_vs_T} plots $\chi(0)$ versus $T$, illustrating exponential growth at low $T$ but scaled down by $\sim 1/k$. These trends underscore how periodicity dilutes the magnetic response, interpolating toward isolated impurities for large $k$.

\section{Correlations}
\label{Correlations}

In statistical mechanics, correlation functions provide crucial insights into the spatial organization of spins, quantifying how fluctuations at one site influence another. For the 1D Ising model, these functions are particularly important as they reveal the absence of long-range order at finite temperatures, a hallmark of one-dimensional systems with short-range interactions \cite{Domb, Baxter}. In our periodic impurity model, correlations acquire an additional layer of complexity due to the modulated field, leading to exponentially decaying envelopes modulated by periodic prefactors. This section derives the two-point correlation function using the transfer matrix formalism, defines the correlation length and strengths, and discusses their physical implications. While the uniform case ($k=1$) yields pure exponential decay, higher $k$ introduces novel oscillatory behaviors, emphasizing the role of impurity spacing.

The two-point correlation function is defined statistically as
\begin{equation}
\langle \sigma_i \sigma_j \rangle = \frac{1}{Z} \sum_{\{\sigma\}} \sigma_i \sigma_j \exp(-\beta \mathcal{H}),
\end{equation}
where the sum runs over all spin configurations, weighted by the Boltzmann factor.

In the matrix formulation, this can be expressed by inserting the spin operator $S = \begin{pmatrix} 1 & 0 \\ 0 & -1 \end{pmatrix}$ at the positions of the spins. Here, $\sigma_i$ is at position $\delta_i$ within the $m$-th unit cell, and $\sigma_j$ is at $\delta_j$ within the $n$-th cell ($n \geq m$). Thus, the numerator in the expression for the correlation is
\begin{equation}
\langle \sigma_i \sigma_j \rangle  = \frac{1}{Z} \mathrm{Tr} \left[ (P_k)^{m-1} A_{\delta_i} (P_k)^{n-m-1} A_{\delta_j} (P_k)^{M-n} \right],
\end{equation}
where $A_{\delta}$ represents the transfer matrix of a unit cell with the spin operator inserted at position $\delta$. Specifically,

\begin{equation}
A_\delta = D(\beta h/2) T_0^{\delta-1} S T_0^{k - \delta + 1} D(\beta h/2),
\end{equation}

Using the spectral decomposition of $P_k$,
\begin{equation}
P_k = \lambda_k^+ |R_+\rangle \langle L_+| + \lambda_k^- |R_-\rangle \langle L_-|,
\end{equation}
where $|R_\pm\rangle$ and $\langle L_\pm|$ are the right and left eigenvectors, we insert this into the trace expression. This yields four terms:
\begin{align}
\langle \sigma_i \sigma_j \rangle Z &= (\lambda_k^+)^{M-2} \langle L_+ | A_{\delta_i} | R_+ \rangle \langle L_+ | A_{\delta_j} | R_+ \rangle\\
&+ (\lambda_k^+)^{m-1} (\lambda_k^-)^{n-m-1} (\lambda_k^+)^{M-n} \langle L_+ | A_{\delta_i} | R_- \rangle \langle L_- | A_{\delta_j} | R_+ \rangle \nonumber \\
&+ (\lambda_k^-)^{m-1} (\lambda_k^+)^{n-m-1} (\lambda_k^-)^{M-n} \langle L_- | A_{\delta_i} | R_+ \rangle \langle L_+ | A_{\delta_j} | R_- \rangle \\
&+ (\lambda_k^-)^{M-2} \langle L_- | A_{\delta_i} | R_- \rangle \langle L_- | A_{\delta_j} | R_- \rangle
\end{align}

In the thermodynamic limit ($M \to \infty$), the terms proportional to $(\lambda_k^- / \lambda_k^+)^M$ and similar vanishing ratios are neglected, leaving
\begin{equation}
\langle \sigma_i \sigma_j \rangle = \frac{\langle L_+ | A_{\delta_i} | R_+ \rangle \langle L_+ | A_{\delta_j} | R_+ \rangle}{(\lambda_k^+)^2} + \left( \frac{\lambda_k^-}{\lambda_k^+} \right)^{n-m-1} \frac{\langle L_+ | A_{\delta_i} | R_- \rangle \langle L_- | A_{\delta_j} | R_+ \rangle}{(\lambda_k^+)^2}.
\end{equation}
The first term represents the disconnected part $\langle \sigma_i \rangle \langle \sigma_j \rangle$, while the second is the connected correlation, defining the correlation strength
\begin{equation}
A_{ij} = \frac{\langle L_+ | A_{\delta_i} | R_- \rangle \langle L_- | A_{\delta_j} | R_+ \rangle}{(\lambda_k^+)^2}.
\end{equation}
Thus, the connected correlation is $A_{ij} \left( \frac{\lambda_k^-}{\lambda_k^+} \right)^{n-m-1}$.

With the distance $r$ between spins and for big n and m, $(n-m-1) \sim n-m = \lfloor r/k \rfloor$, the correlation decays as
\begin{equation}
G(r) \sim A_{ij} \left( \frac{\lambda_k^-}{\lambda_k^+} \right)^{\lfloor r/k \rfloor}.
\end{equation}
The correlation length is then
\begin{equation}
\xi = \frac{k}{\ln(\lambda_k^+ / \lambda_k^-)},
\end{equation}
measuring the scale over which correlations persist.

\subsection{Analysis of the Correlation Length}

\begin{figure}[h!]
\centering
\includegraphics[width=0.8\textwidth]{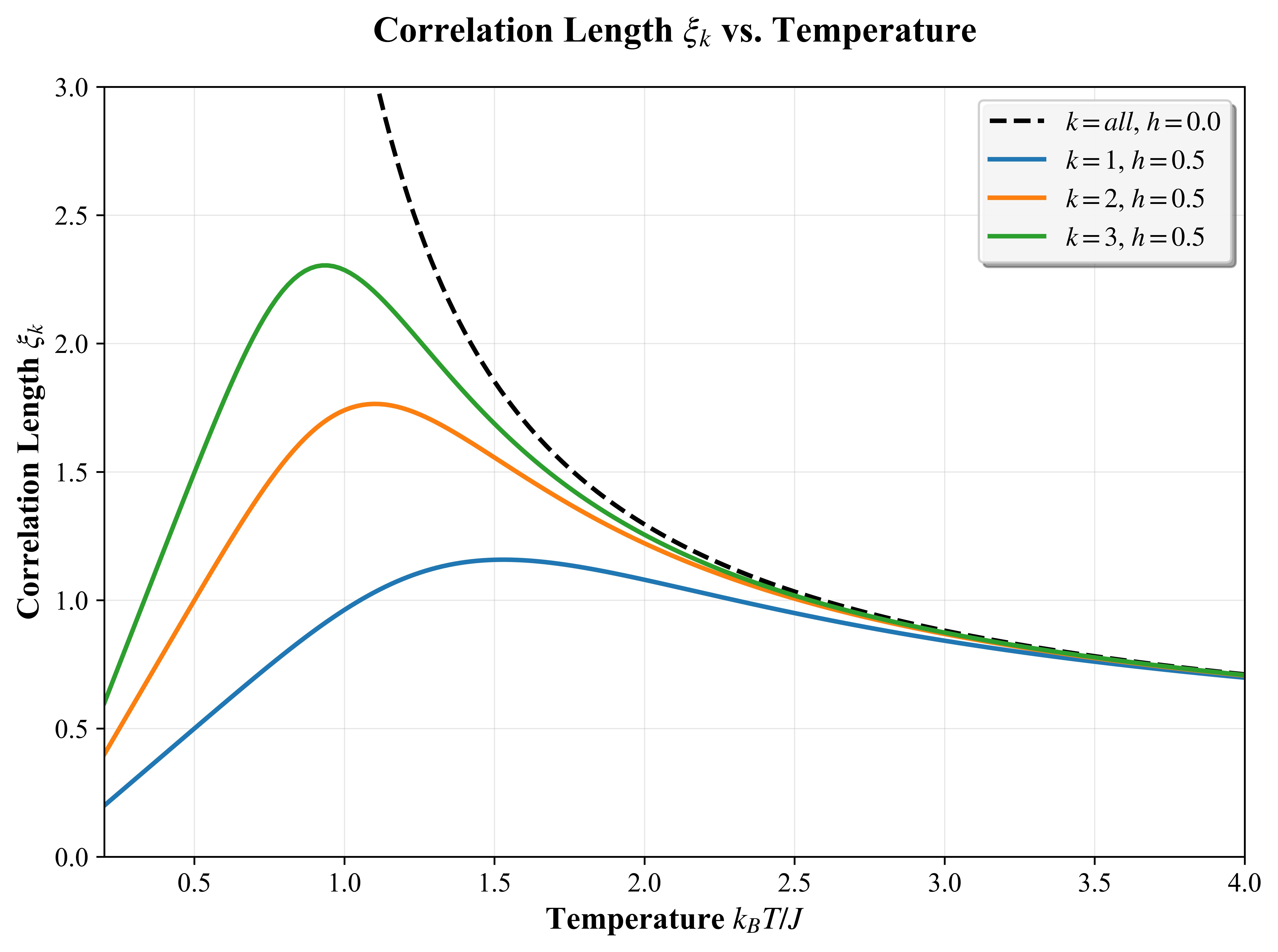} 
\caption{Correlation length $\xi_k$ vs. Temperature ($T$ in $J/k_B$) for $h=0.5J$. This shows the characteristic peak for $h>0$.}
\label{fig:xi_vs_T}
\end{figure}

The exponential decay of $G_c(i,j)$ is characterized by the correlation length $\xi_k$. From the derivation, we identify it as:
\begin{equation}
\xi_k = \frac{k}{\ln(\lambda_k^+ / \lambda_k^-)}.
\end{equation}
The behavior of $\xi_k$ reveals two distinct physical regimes, as shown in Figure~\ref{fig:xi_vs_T}. In the absence of an external field ($h=0$), the correlation length diverges as $T \to 0$, consistent with the system ordering into its ground state. As temperature increases, thermal fluctuations disrupt this order, and $\xi_k$ decays monotonically. Notably, this zero-field correlation length is independent of $k$.

For any non-zero field ($h>0$), the behavior is different. The field creates a spectral gap, and $\xi_k$ is finite at all temperatures. At low $T$, the field freezes the system, suppressing fluctuations and sending $\xi_k \to 0$. As $T$ increases, thermal energy activates fluctuations, causing $\xi_k$ to grow to a maximum, before thermal disorder dominates at high $T$ and $\xi_k$ decays again. This non-monotonic behavior is a classic signature of a gapped 1D system. As seen in Figure~\ref{fig:xi_vs_T}, this peak is more pronounced and occurs at a lower temperature for larger $k$.

\begin{figure}[h!]
\centering
\includegraphics[width=0.8\textwidth]{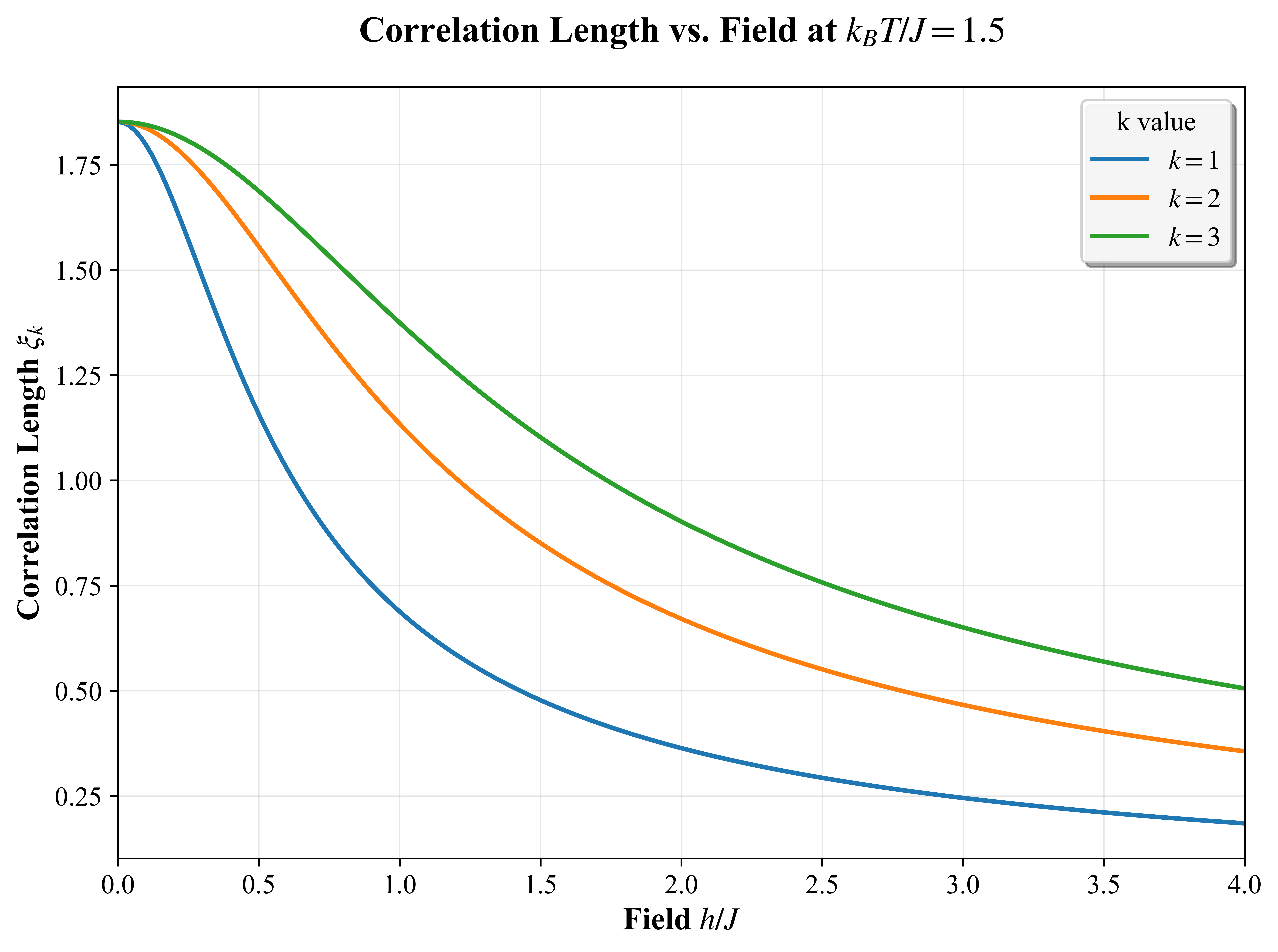}
\caption{Correlation length $\xi_k$ vs. Field Strength ($h/J$) at $T=1.5 J/k_B$. For $h>0$, correlations persist over longer distances for larger $k$.}
\label{fig:xi_vs_h}
\end{figure}

The field's stiffening effect is also visible in Figure \ref{fig:xi_vs_h}. At any fixed temperature, increasing the field strength suppresses fluctuations, causing $\xi_k$ to decay monotonically. Interestingly, for a non-zero field, the correlation length is larger for larger $k$.

\subsection{Analysis of Correlation Strength and Anisotropy}

The prefactor $A_{ij}$ quantifies the strength of the correlation between specific sites. From our derivation, it is defined as:
\begin{equation}
A_{ij} = \frac{\langle L_+ | A_{\delta_i} | R_- \rangle \langle L_- | A_{\delta_j} | R_+ \rangle}{(\lambda_k^+)^2}.
\end{equation}
This prefactor depends strongly on the intra-cell positions $\delta_i$ and $\delta_j$, a phenomenon known as correlation anisotropy.

\begin{figure}[h!]
\centering
\includegraphics[width=1.0\textwidth]{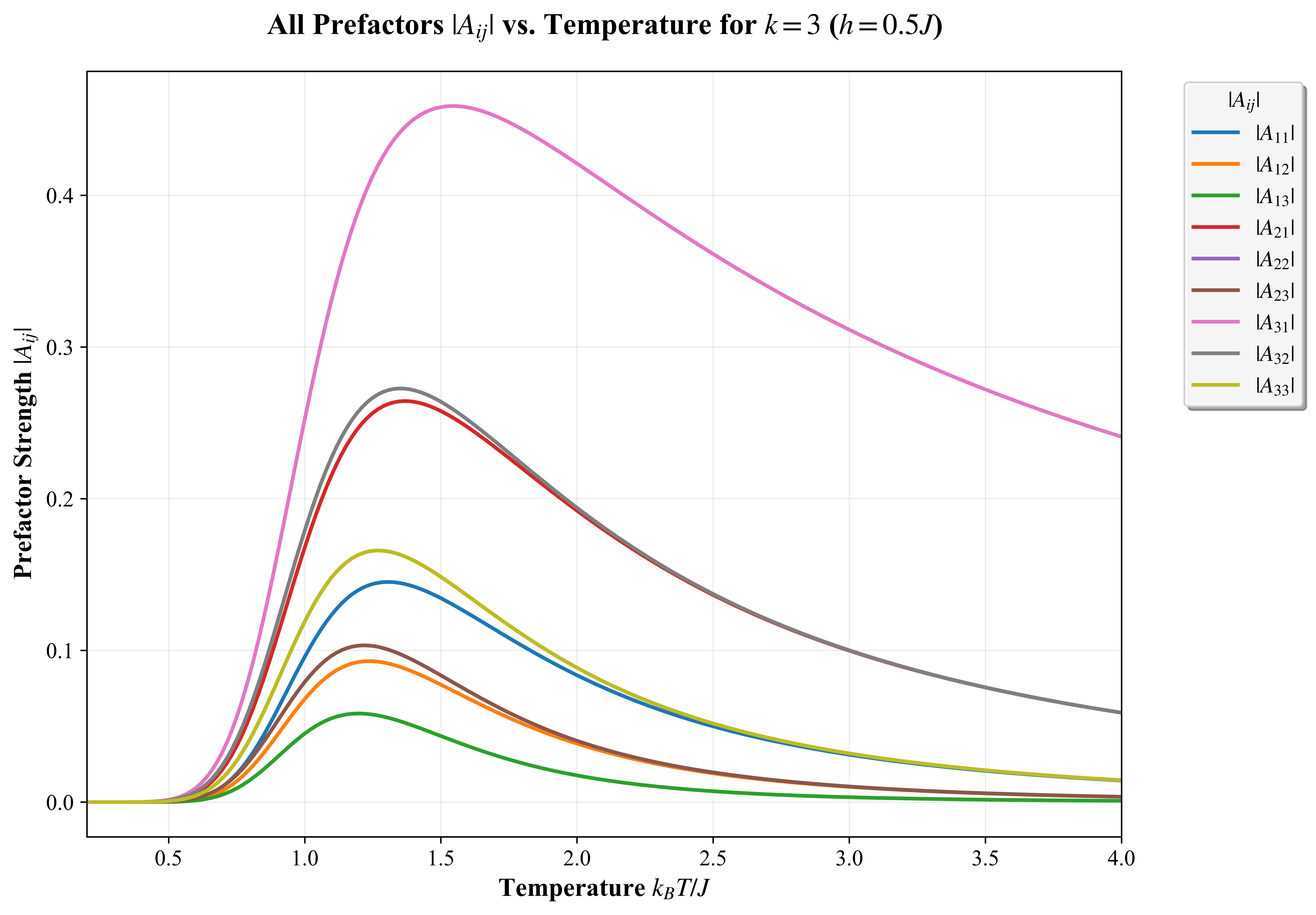}
\caption{The full matrix of correlation strengths $|A_{ij}|$ vs. Temperature for $k=3$, at fixed field strengths. All prefactors exhibit a peak, but their magnitudes vary significantly, demonstrating strong anisotropy.}
\label{fig:prefactors_vs_T}
\end{figure}

The physical behavior of all prefactors, shown for $k=3$ in Figure~\ref{fig:prefactors_vs_T}, mirrors that of the correlation length. They peak at an intermediate temperature, confirming that the correlation strength is maximized in the regime where thermal fluctuations are present but not yet dominant. Similarly, all prefactors decay monotonically as the field $h$ increases, as the field suppresses the fluctuations they describe. The prefactor $A_{ij}$ is inherently asymmetric because the operator $A_\delta$ that measures the spin at a non-impurity site is not symmetric.

The most important finding is the strong anisotropy. As seen in Figure~\ref{fig:prefactors_vs_T}, the magnitudes of the prefactors vary significantly. For $k=3$, the prefactors $|A_{21}|$, $|A_{31}|$, and $|A_{32}|$ are clearly dominant over the "diagonal" correlations like $|A_{11}|$ and $|A_{33}|$. This reveals the microscopic path of correlations: a fluctuation at an impurity site ($\delta=1$) is not most strongly correlated with the next impurity site, but rather with fluctuations at the non-field sites immediately preceding the next impurity. This complex, non-local structure is a direct consequence of the periodic modulation.

\section{Conclusion}
\label{sec:conclusion}

We have presented a complete and exact solution for the one-dimensional Ising model with a periodically applied impurity field of period $k$. By employing a bond-symmetric transfer matrix formalism, we derived the exact eigenvalues and, from them, the general expressions for the system's thermodynamics and correlation functions.

Our analysis highlights several key physical features. The periodic impurity field, as a tunable parameter, systematically suppresses the system's overall magnetic response, as seen in both the magnetization and the zero-field susceptibility. The susceptibility, $\chi_k(0)$, scales as $1/k$ in the large-$k$ dilute limit, confirming physical intuition.

The correlation properties reveal a richer structure. We have shown that the zero-field correlation length $\xi_k(0)$ is independent of $k$, indicating that the intrinsic scale of thermal fluctuations is not affected by the periodic placement of the field sites. However, in a non-zero field, the system becomes gapped, and $\xi_k$ exhibits a non-monotonic dependence on temperature, a hallmark of 1D systems with a finite excitation gap.

Furthermore, our analysis of the correlation strength prefactors $A_{ij}$ demonstrates a strong, position-dependent anisotropy. The dominant correlations are not necessarily between the impurity sites themselves, but often involve non-field sites, revealing a complex, non-local pathway for spin fluctuations. This work provides a rigorous and complete benchmark solution, offering a solid foundation for understanding the effects of periodic modulation in low-dimensional magnetic systems.

\end{document}